\documentclass[prc,showpacs,showkeys,superscriptaddress,twocolumn]{revtex4}
\usepackage{graphicx,epstopdf}
\usepackage{color} 

\begin{document}
\itemindent-3mm
\title{Reply to Comment on: Non-perturbative finite T broadening of the rho
  meson\\
 and dilepton emission in heavy ion-collisions\\ 
 } 

\author{J\"org Ruppert} 
\affiliation{Department of Physics, McGill University, Ernest Rutherford Building, 3600
Rue University, Montreal, QC, Canada, H3A 2T8}
\author{Thorsten Renk}
\affiliation{Department of Physics, PO Box 35 FIN-40014, University of Jyv\"askyl\"a, Finland}
\affiliation{Helsinki Institut of Physics, PO Box 64 FIN-00014, University of Helsinki, Finland}

\begin{abstract}
  {
Nota bene: the numerical calculation underlying Version 1 of this  reply and the original work Phys.Rev. C71:064903,2005 contains a basic numerical error (wrong factor in the self-energy formulas of Phys. Rev.C71:064903,2005). This renders the numerical results presented there and in Version 1 of the reply invalid and enforces a careful reinvestigation of Phys. Rev.C71:064903,2005 and the reply. Calculations for an erratum of Phys. Rev. C71:064903,2005 are in progress.
  } 
\end{abstract}

\date{\today}
\pacs{25.75.-q,11.10.Wx}
\keywords{Vector mesons, dileptons, self-consistent approximations} 
\maketitle
Corrected version of the manuscript to be submitted.

\end{document}